\documentclass[fleqn,usenatbib]{mnras}
\usepackage{newtxtext,newtxmath}

\usepackage[T1]{fontenc}
\usepackage{ae,aecompl}
\usepackage{comment}

\usepackage{graphicx}	
\usepackage{amsmath}	
\usepackage{amssymb}	

\newcommand{\frb}{FRB~180924}
\newcommand{\frbb}{FRB~190523}
\newcommand{\solar}{\(\textup{M}_\odot\)}

\title[Constraining the origin of localized FRBs]{Constraining a neutron star merger origin for localized fast radio bursts}

\author[K. Gourdji et al.]{
K. Gourdji,$^{1}$\thanks{E-mail: \href{mailto:k.gourdji@uva.nl}{k.gourdji@uva.nl} (KG)}
A. Rowlinson,$^{1,2}$
R. A. M. J. Wijers,$^{1}$
A. Goldstein$^{3}$
\\
$^{1}$Anton Pannekoek Institute for Astronomy, University of Amsterdam,S cience Park 904, 1098 XH Amsterdam, The Netherlands\\
$^{2}$ASTRON, Netherlands Institute for Radio Astronomy, Oude Hoogeveensedijk 4, 7991 PD Dwingeloo, The Netherlands\\
$^{3}$Science and Technology Institute, Universities Space Research Association, Huntsville, AL 35805, USA
}

\date{Accepted XXX. Received YYY; in original form ZZZ}

\pubyear{2015}
\begin{document}
\label{firstpage}
\pagerange{\pageref{firstpage}--\pageref{lastpage}}
\maketitle

\begin{abstract}
What the progenitors of fast radio bursts (FRBs) are, and whether there are multiple types of progenitors are open questions. The advent of localized FRBs with host galaxy redshifts allows the various emission models to be directly tested for the first time. Given the recent localizations of two non-repeating FRBs (\frb\ and \frbb), we discuss a selection of FRB emission models and demonstrate how we can place constraints on key model parameters like the magnetic field strength and age of the putative FRB-emitting neutron star. In particular, we focus on models related to compact binary merger events involving at least one neutron star, motivated by commonalities between the host galaxies of the FRBs and the hosts of such merger events/short gamma-ray bursts (SGRBs). We rule out the possibility that either FRB was produced during the final inspiral stage of a merging binary system. Where possible, we predict the light curve of electromagnetic emission associated with a given model and use it to recommend multi-wavelength follow-up strategies that may help confirm or rule out models for future FRBs.  In addition, we conduct a targeted sub-threshold search in \textit{Fermi} Gamma-ray Burst Monitor data for potential SGRB candidates associated with either FRB, and show what a non-detection means for relevant models. The methodology presented in this study may be easily applied to future localized FRBs, and adapted to sources with possibly core-collapse supernova progenitors, to help constrain potential models for the FRB population at large.
\end{abstract}




\section{Introduction}

Fast radio bursts (FRBs) are bright, extragalactic, \mbox{(sub-)}millisecond duration radio flashes of unknown origin \citep{lorimer2007bright,thornton2013population}. Most FRBs are observed to be single events, despite many hours of follow-up observations \citep{petroff2015repeatability}. Lack of repetition challenges our ability to localize them precisely, which would provide vital clues in understanding their elusive progenitors. 
Even repeating FRBs can be challenging to localize given their sporadic activity \citep{CHIMErepeat19}. The precise localization of a large sample of repeating and non-repeating FRB sources is required to address the central question of whether there are multiple origins and, as a by-product, whether or not all FRB sources are intrinsically repeaters.

There is a long list of FRB origin theories and an overview is provided in \citet{platts18}\footnote{\url{https://frbtheorycat.org}}. Most viable repeating FRB models, though, involve a neutron star (NS) that is either magnetically or rotationally powered.  
The observation of repeat bursts from about 20 percent of known FRB sources \citep{CHIMErepeaters19,Fonseca2020} raises the possibility of multiple FRB origins. Alternatively, all FRBs may repeat but their observable repeat may vary from one source to another depending on their environment or intrinsic bursting rate. Indeed, the high FRB rate compared to the rate of possible progenitors likely implies that the majority of FRB sources repeat \citep{RaviRates}. In any case, the contrast between the environments of repeating and (observed) non-repeating sources lends support to the possibility of multiple progenitors. Additionally, the characteristic of downward drifting sub-bursts in frequency, revealed in some repeat bursts of most repeating sources, has yet to be observed in a one-off FRB \citep{hessels18,CHIMErepeaters19,Fonseca2020}. This burst morphology may serve as another diagnostic to distinguish between (observed) non-repeating and repeating FRB sources.

FRB 121102 \citep{spitler2014fast}, the most active repeating source \citep{spitler2016repeating}, was the first FRB to be precisely localized, thanks to very long baseline interferometry \citep[][]{chatterjee2017direct,Marcote17}. It was associated with a low-metallicity dwarf galaxy at redshift $z=0.19$ \citep{tendulkar2017host}, in a region of active star formation \citep{bassa2017frb,kokubo17}, and coincident with a persistent radio source ($1.8\times10^{29}$\,erg\,s$^{-1}$\,Hz$^{-1}$, \citealp{chatterjee2017direct,Marcote17}). Furthermore, the bursts exhibit an enormous and variable rotation measure (RM $\sim10^{5}$rad\,m$^{-2}$), placing them in an extreme magneto-ionic environment \citep{michilli2018}. The host galaxy of FRB~121102 shares similar properties with the environments of long gamma-ray bursts (LGRBs) and type Ibc super-luminous supernovae \citep{tendulkar2017host,Marcote17}, which have massive star progenitors. A related interpretation is that the persistent radio source is a nebula powered by a magnetar, supplying a highly magnetized plasma \citep[e.g.][]{Murase16,Beloborodov17,CaoMagnetar,Metzger17,Nicholl17}. Alternatively, the large RM and persistent radio emission may be due to an AGN in the vicinity of the bursting source \citep[e.g.][]{Marcote17,michilli2018,zhang18comb}.

A second repeater, FRB~180916.J0158+65, was localized by \citet{Marcote2020} to an outer arm of a nearby spiral galaxy in a star-forming region. Unlike FRB~121102, there is no comparably bright persistent radio source nor significant Faraday rotation. This result indicates that sources of repeat FRBs may reside in a variety of galaxy types and environments.

In 2019 August, the localizations of two non-repeating FRBs were reported. FRB~180924 was localized to milliarcsecond precision and, unlike FRB~121102 and FRB~180916.J0158+65, repeat bursts have not been detected from this source in approximately 11 hours of follow-up observations conducted over two separate observing sessions separated by two weeks \citep{Bannister19}. The host galaxy of FRB~180924 is markedly different from the environment of FRB~121102. Namely, the host is a spiral galaxy ($z=0.32$) with limited star formation, there is no persistent source of radio emission above $7\times10^{28}$\,erg\,s$^{-1}$\,Hz$^{-1}$, and the burst has a negligible RM of $14$\,rad\,m$^{-2}$ \citep{Bannister19}. The other source, \frbb, was localized with arcsecond accuracy to a massive galaxy at $z=0.66$ with limited star formation activity ($<1.3$\,\solar\,yr$^{-1}$) \citep{Ravi19loc}. There is no associated constant radio emission greater than $7\times10^{30}$\,erg\,s$^{-1}$\,Hz$^{-1}$. Polarimetric information is not available for this source. No repeat bursts were observed in 78 hours of follow-up observations conducted within a span of 54 days. The environments of these localized sources both have low star formation rates, which contrasts the active star formation regions associated with the only two localized repeating sources\footnote{A third source of a singular FRB was more recently localized by \citet{Prochaska19}, bringing the total of localized sources to five (2 repeating and 3 non-repeating).}.

Interestingly, both  \frb\ and \frbb\ were emitted in the outskirts of their host galaxy. The limited star formation (pointing to an older stellar population) and positional offset from their hosts are consistent with a neutron star merger origin (binary neutron star, BNS, or black hole neutron star, BHNS). \cite{margalit19} and \citet{Wang2020} show that the environments of \frb\ and \frbb\ are consistent with the population of SGRBs, which are produced during BNS and possibly BHNS mergers. 
Comparisons between the rates of FRBs and neutron star mergers show that only a fraction of non-repeating FRBs could be produced via BNS or BHNS mergers, but that if most or all FRBs repeat on sufficiently long timescales, the rates are adequate for FRBs to emanate from neutron stars born out of BNS mergers \citep{Cao18,RaviRates,margalit19,Wang2020}. 

In this paper, we explore the scenario in which \frb\ and \frbb\ are associated with a compact binary merger involving at least one neutron star. We consider six models (some capable of producing repeat bursts) within the BNS and BHNS merger scenarios and place limits on key parameters within each model using the observed properties of both FRBs. Where applicable, we demonstrate the value of multi-wavelength data sets. In addition, we perform targeted searches for associated SGRBs in \textit{Fermi} Gamma-ray Burst Monitor~\citep[GBM;][]{Meegan2009} data. We emphasize that most of the models we present can be adapted to FRBs related to core-collapse supernovae. In \textsection\ref{models}, we describe the models being considered and in \textsection\ref{sect:GRB search} we demonstrate our SGRB search. We present and discuss our results for each model in \textsection\ref{sect:disc} and draw our main conclusions in \textsection\ref{conclusion}. 

\section{Relevant FRB models}
\label{models}
In this section, we provide an overview of the FRB models we have chosen to examine using the measured parameters of \frb\ and \frbb. The models are organized by the merger stage in which they are expected to occur. We consistently use the following definition for FRB luminosity, unless stated otherwise:
\begin{equation}
\centering
 L_{\text{FRB}} = \Omega F_\nu \Delta \nu D^2\,,
\label{eq:luminosity FRB}
\end{equation}
where $\Omega$ is the solid angle illuminated by the beam of emission ($0<\Omega\leq4\pi$), $F_\nu$ is the flux measured across the observing frequency bandwidth, $\Delta \nu$, and $D$ is the luminosity distance. We implicitly assume a flat spectrum across the observing bandwidth. In using $\Delta \nu$ as opposed to, for example, the observing frequency, we make no assumption about the breadth of the intrinsic spectrum of emission. However, we are likely underestimating the luminosity in this way, since the emission of both FRBs is presumably detectable beyond the observing bandwidth, though to unknown extents \citep[see][]{Gourdji19}. We shall comment on the impact this has on the models in the sections that follow.

\subsection{Pre-merger}
\label{sect:battery}

If at least one neutron star is magnetized in an inspiralling compact binary system, as the companion (BH or NS) moves through the magnetosphere of the charged neutron star, a current may be driven through the magnetic field lines that connect the system, like a battery. This surge accelerates charged particles along the field lines and electromagnetic (EM) emission may be produced.
The total ``battery'' power available for extraction into electromagnetic emission increases as the orbital separation decreases and orbital velocity increases, and so the emission may only be detectable during the final stages of the inspiral. The emission is expected to peak at the point of contact (or point of tidal disruption) of the binary system. Either hemisphere of the conducting companion forms a closed circuit with either magnetic pole of the primary neutron star. The voltage induced along the magnetic field lines can be expressed as \citep{McwilliamsLevin11, Piro12,Dorazio16} 
\begin{equation}
    \oint \Big(\frac{\textbf{v}}{c}\times\textbf{B}\Big)\cdot d\textbf{l}\,,
\end{equation}
where $\textbf{B}$ is the magnetic field vector, $d\textbf{l}$ is the segment that contributes to the electromotive force, $\textbf{v}$ is the relative orbital velocity of the conducting companion (neglecting the magnetized neutron star's rotation) and $c$ is the speed of light. Due to the dot product, only those segment components parallel to the induced electric field contribute ($2R$).  The potential difference, $V(r)$, across one hemisphere of the conducting companion is then
\begin{equation}
    V(r) = 2R\frac{v}{c}B\Big(\frac{R_{\text{NS}}}{r}\Big)^3 \,,
    \label{eq:voltage}
\end{equation}
where $r$ is the orbital separation, $R$ is the radius of the conducting companion, $R_{\text{NS}}$ is the radius of the primary neutron star, and the last term comes from the fact that the strength of the magnetic field drops off as the distance cubed. The total battery power from both hemispheres is then
\begin{equation}
    P=2\frac{V^2}{\mathcal{R}}\,,
    \label{eq:power}
\end{equation}
where $\mathcal{R}$ is the resistance of the system. The resistance across the horizon of a black hole is $\frac{4\pi}{c}$ \citep[impedance of free space,][]{znajek78}. The resistance across a neutron star's magnetosphere is less obvious, but can at most reduce the total power by approximately one half, so we therefore neglect it from our analysis for simplicity (see \citealp{McwilliamsLevin11} for discussion). 

In a BHNS system, the BH (out to horizon radius $R_\text{H}=\frac{2GM}{c^2}$) is the conductor that induces the electromagnetic force along the NS's field lines as it orbits through them \citep[e.g.][]{Mingarelli15}. The total battery power available for conversion into radio emission increases as the orbital separation decreases. We consider the maximal energy case, where the closest orbital separation is the photon sphere radius, $r=\frac{3GM}{c^2}$ (note that this is $\frac{GM}{c^2}$ for a spinning BH). The resulting power is then (combining equations \ref{eq:voltage} and \ref{eq:power})
\begin{equation}
   P = \frac{8c}{729\pi}\Big(\frac{GM}{c^2}\Big)^{-4}B^2 R_{\text{NS}}^6\Big(\frac{v}{c}\Big)^2.
   \label{eq:battery power}
\end{equation}
We take $\frac{v}{c}\approx1$ for simplicity, take the mass of the black hole to be 10\,\(\textup{M}_\odot\), and $R_\text{NS}=10$\,km. Solving equation \ref{eq:battery power}, we end up with battery radio luminosity
\begin{equation}
    L = 2\times10^{45}\Big(\frac{B}{10^{13}\,\text{G}}\Big)^2\epsilon_r\, \text{erg s$^{-1}$}\,,
    \label{eq:batteryBH}
\end{equation}
where some fraction $\epsilon_r$ of the total battery power available is converted into radio emission, depending on the method of energy extraction. We caution that there are caveats to using both smaller and larger BH masses, as the equations may no longer be appropriate (NS plunging into the BH versus tidal disruption). These are addressed in \citet{McwilliamsLevin11} and \citet{Dorazio16}.  

For a binary neutron star system, equation \ref{eq:voltage} is used with $v=\omega r$, where $\omega=\sqrt{\frac{2GM}{r^3}}$ is the orbital frequency (we neglect contribution to $V$ from the neutron star spin) \citep{MetzgerZivancev16}. Using $\mathcal{R}=\frac{4\pi}{c}$ for the primary neutron star's magnetosphere, and minimizing the orbital separation to the point of Roche contact \citep[$r=2.6R$,][]{Eggleton83}, the total power available is then
\begin{equation}
    P = \frac{4GMB^2R^8}{cr^7\pi}
\label{eq:battery_BNS_power}
\end{equation}
Setting $R=10$\,km and $M=1.4$\,\solar, the radio luminosity can then be expressed as:
\begin{equation}
    L = 1\times10^{45}\Big(\frac{B}{10^{13}\,\text{G}}\Big)^2\epsilon_r\, \text{erg s$^{-1}$}\,.
\label{eq:battery_BNS}
\end{equation}
This idea of radio emission from inspiralling BNS systems has also been considered in \citet{HansenLyutikov2001} but with slight differences (in particular a more complex treatment of the electrodynamics) that amount to a larger derived maximum luminosity by almost an order of magnitude (also see \citealp{Lyutikov13}, equation 12). The model is revisited in \citet{Lyutikov19} with two magnetized neutron stars.  \citet{Piro12} expanded on the BNS battery system, demonstrating the dissipation energy available as a function of time, and paying particular attention to the resistance of the circuit.

\subsection{During a SGRB}
\label{sect:GRB}
If a GRB jet that is powered by a Poynting flux dominated wind is launched following a merger, radio emission may be generated at the shock front and detected as an FRB, if the radio waves can escape through it \citep{UsovKatz2000}. This mechanism requires a highly magnetized wind, which is assumed to come from a rapidly rotating and highly magnetized central engine (a neutron star or an accretion disk around a black hole). The magnetic field of the shock front between the wind and ambient medium, in the rest frame of the wind, is \citep{UsovKatz2000}:
\begin{equation}
    B_0 = \epsilon_{B}^{\frac{1}{2}}BR^3c^{-2}\Big(\frac{2\pi}{P_0}\Big)^{2}Q^{-\frac{1}{3}}n^{\frac{1}{3}}\Gamma^{-\frac{1}{3}}\,,
\end{equation}
where $\epsilon_B$ is the fraction of wind energy contained in the magnetic field, $R$ is the radius of the compact object, $P_0$ is its initial spin period, $B$ is the surface magnetic field of the disk or NS, $n$ is the density of the ambient medium, $Q$ is the kinetic energy of the wind assuming spherical outflow, and $\Gamma$ is the Lorentz factor. We shall assume standard values $B = 10^{14-16}$\,G, $R=10^6$\,cm, $P_0 = 1-10$\,ms, $Q=10^{53}$\,ergs, $\Gamma=1000$, $n=10^{-2}$\,cm$^{-3}$. The peak radio emission frequency, $\nu_{\text{max}}$, is then the gyration frequency of an electron in a magnetic field $B_0$, which works out to $\nu_{\text{max}} \approx \frac{eB_0}{c m_p}$ \citep[][equation 5]{UsovKatz2000}. In this model, we assume that the radio emission ranges from the gyration frequency to the observing frequency of the FRB. The bolometric radio fluence, $\Phi_r$, is then:
\begin{equation}
    \Phi_r = \frac{\mathcal{F}_{\text{FRB}}}{\nu_{\text{obs}}^{\alpha}(\alpha + 1)}\big(\nu_{\text{obs}}^{\alpha+1}-\nu_{\text{max}}^{\alpha+1}\big)\,
     \label{eq:usov_radio}
\end{equation}
where $\mathcal{F}_{\textup{FRB}}$ is the measured fluence of the FRB, $\nu_{\text{obs}}$ is the observing frequency, and $\alpha$ is the spectral index assumed to be $-1.6$. According to \citet{UsovKatz2000}, the bolometric gamma-ray fluence, $\Phi_{\gamma}$ is related to $\Phi_r$ as
\begin{equation}
    \frac{\Phi_r}{\Phi_{\gamma}} \approx 0.1\epsilon_{B}\,.
    \label{eq:usov_ratio}
\end{equation}
 Combining equations \ref{eq:usov_radio} and \ref{eq:usov_ratio}, one can solve for the expected gamma-ray fluence, $\Phi_\gamma \propto \epsilon_B P_0^{-2}B$. Given that the radio and gamma-ray emission arise from the same region, beaming effects should cancel in equation \ref{eq:usov_ratio}.

\subsection{Post-merger}

\subsubsection{Pulsar-like emission}
\label{sect:pulsar}
If the merger remnant is a neutron star, it may be detectable through pulsar emission from its amplified magnetic field.
A comparison to the energetics of the population of known radio pulsars will quickly reveal a disparity spanning several orders of magnitude relative to the energy of FRBs. Therefore, pulsar giant pulse emission \citep[an observational term referring to pulses with fluence greater than some multiple, typically taken to be 10, of the average,][]{Karuppusamy2010} has often been invoked in an effort to close this gap, in the rotationally powered pulsar model for FRBs. This is because giant pulses offer more freedom in the parameter space available. Specifically, one can say that giant pulses result from increases in efficiency and/or beaming. 
Following the model described in \citet{PshirkovPostnov2010}, it is assumed that the radio luminosity $L$ is equal to some fraction of the energy loss rate of the magnetically driven outflow, $|\dot{E}|$. Therefore, it follows that (using equation \ref{eq:luminosity FRB})
\begin{equation}
\begin{split}
L &= \epsilon_r|\dot{E}|\,,\hspace{1cm}\\ 
F_{\nu} &= \frac{\epsilon_r|\dot{E}|}{\Omega \Delta \nu D^2}\,,
\end{split}
\label{eq:pulsar}
\end{equation}
for the predicted emission.
In equation \ref{eq:pulsar}, $\epsilon_r$ encapsulates all unknowns related to the emission mechanism and simply says that some fraction of the dipole energy is converted into the observed radio emission. The standard pulsar spin-down equation is \citep{pulsarhandbook}
\begin{equation}
\dot{E}=\frac{16\pi^4}{3}\frac{B^2R^6}{P^4c^3}\,,
\label{eq:spindown}
\end{equation}
where $P$ is the spin period, $B$ is the magnetic field at the surface of the neutron star and $c$ is the speed of light. The angle between the magnetic moment and the spin axis is a source of uncertainty and depends on the physics of the NS magnetic field and EOS, but is thought to be near zero at the time of birth, and is expected to increase with time \citep[e.g.][]{Dallosso09}. We have therefore assumed a fiducial value of $30$\textdegree, and note that a range from 1\textdegree\ (nearly aligned spin and magnetic axes) to 90\textdegree\ (orthogonal spin and magnetic axes) corresponds to about an order of magnitude difference for the derived NS magnetic field. Plugging $\dot{E}$ from equation \ref{eq:spindown} into equation \ref{eq:pulsar} and solving for $B$, we find that its dependence on the three most uncertain quantities is $B\propto \Omega^{\frac{1}{2}}P^2\epsilon_r^{-\frac{1}{2}}$. 
\subsubsection{Flaring magnetar}
\label{sect:magnetar}
An alternative to rotational energy extraction is magnetically powered neutron star emission. A popular subclass is the flaring magnetar \citep{lyubarsky2014maser,Beloborodov17,metzger19,Beloborodov19}. In this scenario, giant flares,
caused by instabilities in the magnetosphere, shock the plasma surrounding the magnetar to produce maser emission detectable at radio frequencies. In order to establish the relevant parameters, we outline and build on the model presented in \citet{lyubarsky2014maser}. The various models diverge at different steps, but possibly the most important differences lie in the nature of the upstream/shocked material, which we address in \textsection\ref{disc:mag}.

The magnetar flares start in the form of  magneto-hydrodynamic waves (Alfv\'{e}n waves) that propagate in the magnetosphere, sweeping up field lines to form a pulse that travels through the magnetar's wind. The wind is composed  of magnetized electron positron plasma and its luminosity is determined by the spin-down luminosity. The wind's end boundary occurs when the wind's bulk pressure is balanced by the pressure confining the wind. There is a termination shock at the radius at which this balance occurs, and a hot wind bubble (like a nebula) consequently forms. When the pulse reaches the termination shock, it meets a discontinuity as the upstream medium suddenly changes from the cold wind to the hot wind/nebula. It blasts the plasma in the nebula outward, generating a forward shock that propagates through the nebula's plasma. The magnetic field of the wind runs perpendicular to the pulse and the shock is mediated by that field. The gyration of the shocked particles creates an unstable synchrotron maser, that produces low-frequency emission, a fraction of which ($\eta$) manages to escape thermalization through the upstream unshocked plasma. For a burst of duration $\Delta t$, the isotropic energy of the escaped emission is \citep{lyubarsky2014maser}:
\begin{equation}
    E_{\text{iso}} = \frac{\eta B^2R^2nm_ec^3b^2\Delta t}{16p\xi}\,,
    \label{eq:magnetar_energy_full}
\end{equation}
where $B$ is the surface magnetic field of the magnetar, $b$ is the fraction of $B$ contained in the magnetic pulse, $R$ is the magnetar radius, $p$ is the pressure of the nebula and $n$ is its particle density, $m_e$ is the electron rest mass, and $\xi$ takes into account the fraction of high energy particles in the shocked plasma that will lose their energy before being able to enter the upstream nebula. The detailed derivation of equation \ref{eq:magnetar_energy_full} is in Appendix \ref{sect:appendix}.

\subsubsection{Curvature radiation}
\label{sect:curvature} 
Apart from maser emission, which was explored in the previous section, particle energy may be dissipated through curvature radiation and detected as an FRB. Models where the FRB is produced within the magnetosphere have the advantage of not having to deal with the potentially critical effects of induced Compton scattering, which lead to losses in photon energy \citep{LuKumar18}. In the model presented by \citet{Kumar17}, particles are accelerated by an electric field parallel to the magnetar's magnetic field lines. Based on this idea, \citet{LuKumar19} show that the FRB luminosity is limited by the parallel electric field, $E_{||}$, which can be at most $5$ per cent of the quantum critical field \citep[$\frac{m^2c^3}{e\hbar}$$\approx 4.4\times10^{13}$\,esu for electrons,][]{StebbinsYoo15}, else the electric field gets shielded by Schwinger pairs. The moving particles will induce a magnetic field, $B_{\text{ind}}$, perpendicular to the field lines. This induced field must not perturb the original magnetic field, $B$, by more than a factor of the beaming angle $\gamma^{-1}$, else coherence is lost. Applying the requirements that i) $E_{||}<2.5\times10^{12}$\,esu and ii) $B_{\text{ind}}<B\gamma^{-1}$, and following \citet{LuKumar19}, the following requirement on $B$ can be set:
\begin{equation}
    B > \frac{(2\pi)^{2/3} L_{\text{iso}}\nu^{2/3}}{E_{||}\rho^{4/3}c^{5/3}} \approx 6 \times 10^{12}\,\text{G}\Bigg(\frac{L_{\text{iso}}}{10^{44}\,\text{erg s}^{-1}}\Bigg)\Bigg(\frac{\nu}{1.4\,\text{GHz}}\Bigg)^{2/3}\,,
    \label{eq:curvature}
\end{equation}
where $\rho$ is the curvature radius, taken to be $10^6$\,cm (the local magnetic field may be too weak at larger values), and $\nu$ is the peak frequency of the emission, taken to be the central observing frequency of the FRB. The spectrum of the predicted emission is broadband. 

\subsubsection{Neutron star collapse}\label{sect:collapse}
Here, we entertain the scenario where the post-merger product is a neutron star that collapses at some point into a black hole. During this process, the neutron star ejects its magnetosphere (according to black hole no-hair theorem) emitting a short duration burst of coherent radio emission \citep{FalckeRezzolla14}. \citet{zhang14} estimate that the total amount of magnetic energy $E_B$ stored in the magnetar's magnetosphere is approximately $\frac{1}{6}B^2R^3$, of which some fraction $\epsilon_r$ is converted into coherent radio emission. The radio luminosity is then
\begin{equation}
L = \epsilon_r\frac{E_B}{\Delta t} = \epsilon_r\frac{B^2R^3}{6\Delta t}\,.
\label{eq:lum_collapse}
\end{equation}
Rearranging equation \ref{eq:lum_collapse} and using equation \ref{eq:luminosity FRB}, we can solve for the magnetic field at the surface of the collapsing merger remnant,
\begin{equation}
    B= \Bigg[\frac{6F_{\nu}\Omega D^2\Delta t\Delta\nu}{\epsilon_r R^3}\Bigg]^{\frac{1}{2}}\,.
    \label{eq:collapse}
\end{equation}

\section{Search for SGRB counterparts}
\label{sect:GRB search}
All FRB models in this study should theoretically have a SGRB counterpart. The expected amount of time elapsed between the SGRB and FRB detections is model dependent and can range from decades (\textsection\ref{disc:mag}) after the SGRB to seconds before the SGRB (\textsection\ref{disc:magnetosphere}). \citet{margalit19} checked archival data for positionally coincident SGRBs that could be associated with \frb. Given the lack of positional accuracy of most GRB detectors, there are naturally several possible associations over the last decades. We perform the same check for \frbb\ using the \textit{Swift}/BAT catalogue\footnote{\url{https://swift.gsfc.nasa.gov/archive/grb_table/}} as it provides by far the best positional accuracy (on average a positional error radius of only $1.6'$). An association was not found, though the instrument's instantaneous field of view is roughly only $15$ per cent of the sky and could therefore conceivably miss a GRB counterpart to an FRB.

The \textit{Fermi} GBM, however, sees about 65 per cent of the sky. Therefore, to test the possibility of detecting a temporally coincident GRB counterpart, we study the data from the \textit{Fermi} GBM during a window of 30 minutes prior to and 1 minute after the FRB detections. In order to find any signal that may be too weak to trigger the detectors on-board the spacecraft, and to set the most stringent flux upper limits in the absence of any such signal, we utilize the GBM targeted sub-threshold search that was developed to search for short GRB counterparts to gravitational-wave signals~\citep{Blackburn2015, Goldstein2019, Hamburg2020}.  The targeted sub-threshold search operates by performing a spectrally and detector-coherent search of the GBM data around a time window of interest, and it has been validated by showing that it can recover GRBs detected by the Swift BAT, but that were at unfavorable arrival geometries or too weak to trigger the onboard detection algorithms~\citep{Kocevski2018}.  In addition to gravitational-wave follow-up~\citep{Burns2019}, the targeted search has been utilized to search for short GRB counterparts to astrophysical neutrinos~\citep[e.g.;][]{IceCube200107A,IceCube200117A} and to other FRBs~\citep{Cunningham2019}.   

\begin{figure}
    \centering
    \includegraphics[scale=0.435]{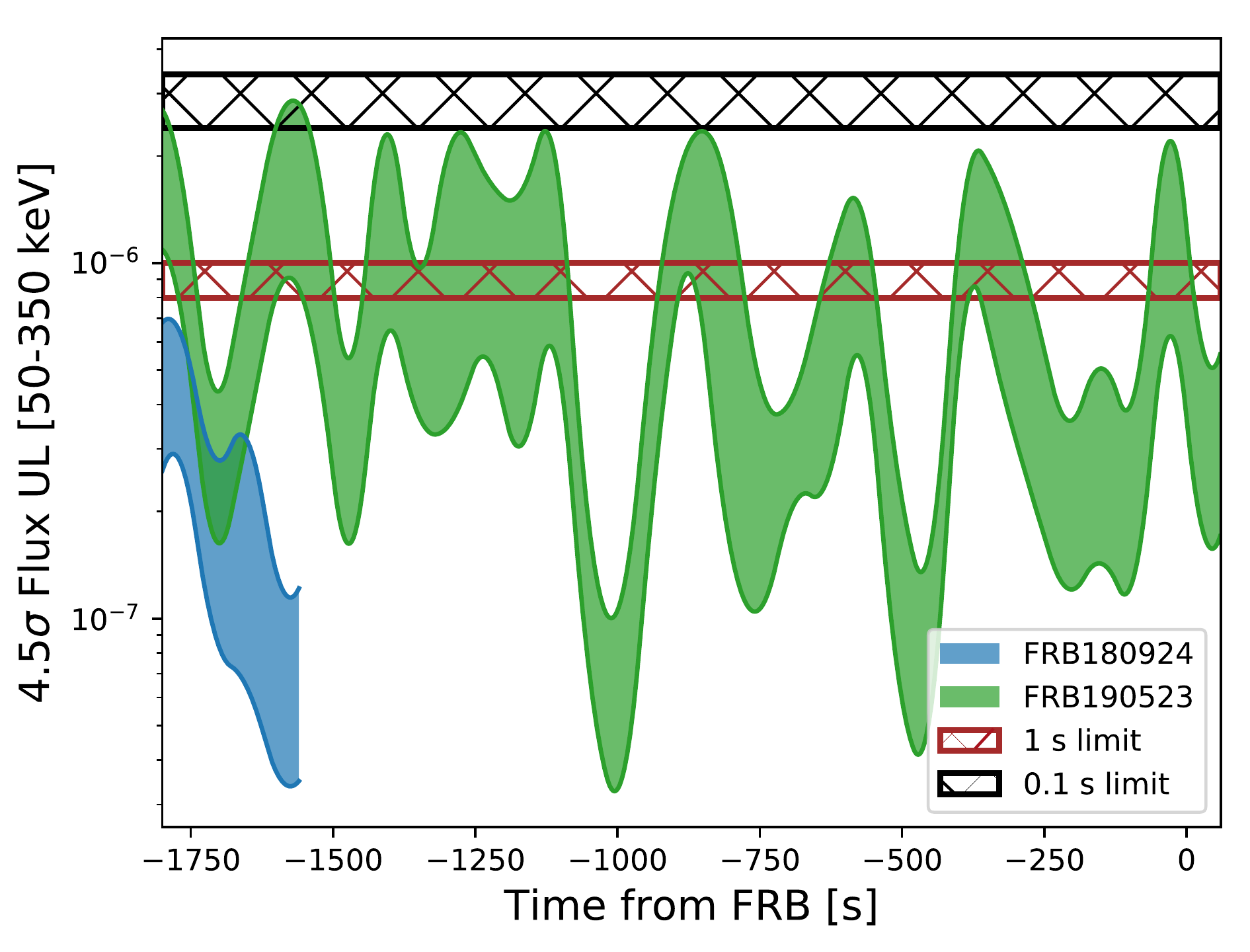}
    \caption{Time-dependent flux upper limits from \textit{Fermi} GBM for a short GRB counterpart to \frb\ and \frbb. The shaded regions correspond to the upper limits for a signal between 0.1 s and 1 s in duration.  The location of \frb\ was only visible to GBM until 26 minutes prior to the FRB detection.  The brown and blacked hatched region show the approximate flux upper limit for a single GBM detector observing an FRB location at $70^\circ$ from the boresight, for a 1 s and 0.1 s duration signal, respectively.  The thickness of the hatched regions denote the variability in this upper limit due to non-stationary background.}
    \label{fig:upperlimits}
\end{figure}

Operating the GBM targeted search during the (-30, +1) minute window around each FRB, we find that while GBM was able to observe the location of \frbb\ during the full window, unfortunately the location of \frb\ was only visible until $\sim26$ minutes prior and then was occulted by the Earth for the remainder of the window. The targeted search did not find any promising candidates, however we can place time-dependent coherent flux upper limits for the known positions of the FRBs, which is shown in Figure~\ref{fig:upperlimits}.  For a signal with duration between 0.1\,s and 1\,s in the 50--350 keV band, the $4.5\sigma$ flux upper limits are typically below $\sim10^{-6}$ erg s$^{-1}$ cm$^2$, with time-dependent variations that span more than an order of magnitude as a result of the spacecraft orbital and pointing motion relative to the source positions. Additionally, for comparison, we estimate the flux upper limit for a single GBM detector observing the FRB position at an angle of $70^\circ$ to the detector boresight, which is a good proxy for a very poor observing scenario for a single GRB scintillation detector. This is done by choosing a few random intervals in a single detector during the 31-minute period considered in the search. For each interval, a local background is fit, a detector response is generated assuming a source angle of $70^\circ$ from the detector boresight, and the spectral amplitude for the same spectrum used in the targeted search is fit to the data above background. From the assumed spectrum and fitted amplitude, we estimate the corresponding flux upper limit. We find this upper limit to be $\sim10^{-6}$ erg s$^{-1}$ cm$^2$ for a 1-s duration signal, and the variance of this upper limit shown in Figure~\ref{fig:upperlimits} is from the range of upper limits calculated from the chosen random intervals. We comment on the results of this SGRB search throughout the following section.
\section{Results and discussion}
\label{sect:disc}

In this section, we use the measured properties of \frb\ and \frbb, as well as the upper limits on an SGRB counterpart from the previous section, to place constraints on the models described in \textsection\ref{models}. In particular, we make use of the FRB properties listed in Table \ref{tab:props} for convenience.

\begin{table}
\centering
\begin{tabular}{lll}
\hline
 Property & \frb & \frbb   \\
 \hline
 $F_{\nu}$ (Jy) & 12.3  & 666.7   \\
 $\Delta t$ (ms)& 1.3 &  0.42 \\
 $L_{\text{iso}}$ (erg s$^{-1}$ Hz$^{-1}$)&  $4.2 \times 10^{34}$ & $1.3\times10^{37}$ \\
 $z$ & 0.3214 & 0.660 \\
 $D_L$ (Gpc) & 1.7 & 4.0 \\
 $\Delta \nu$ (MHz) & 336 & 153\\
 $\nu_{\text{obs}}$ (GHz) & 1.32 & 1.411\\
\end{tabular}
\caption{Relevant FRB properties. $F_{\nu}$: burst flux density; $\Delta t$: burst duration; $L_{\text{iso}}$: isotropic luminosity; $z$: host galaxy redshift; $D_L$:Luminosity distance derived using $z$ and assuming a cosmology with $H_{0}=69.6$\,km\,s$^{-1}$\,Mpc$^{-1}$, $\Omega_{m}=0.286$ and $\Omega_{\lambda}=0.714$ \citep{CosmologyCalc}; $\Delta \nu$: observing bandwidth; $\nu_{\text{obs}}$: observing frequency.}
 \label{tab:props}
\end{table}

\subsection{Magnetospheric interactions}
\label{disc:magnetosphere}
\begin{figure*}
    \centering
    \includegraphics[width=\textwidth]{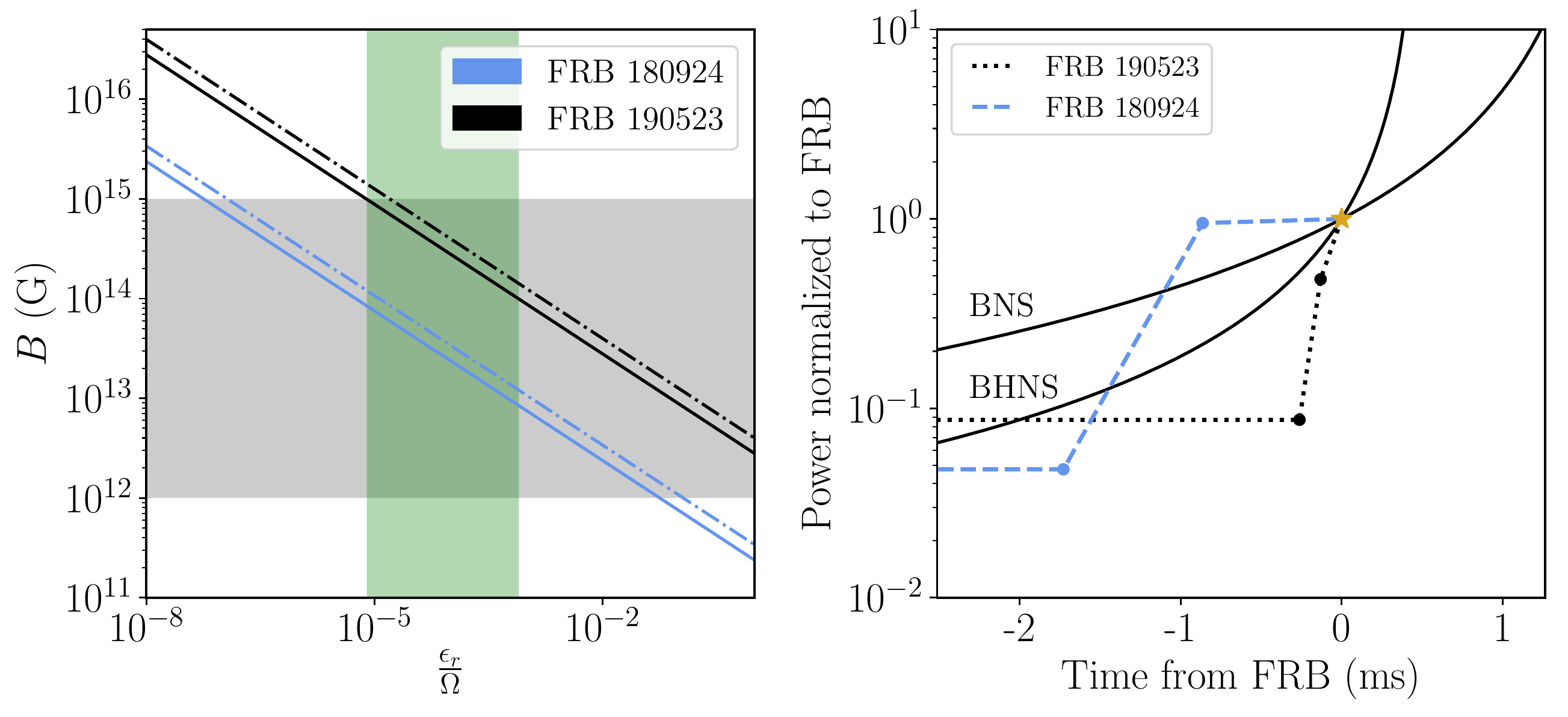}
     \caption{\textit{Left}: Following equations \ref{eq:batteryBH} and \ref{eq:battery_BNS}, derived magnetic field, $B$, of the (primary) neutron star in a BNS and BHNS system, respectively, that produces an FRB through the battery model outlined in \textsection\ref{sect:battery}, as a function of the radio efficiency, $\epsilon_r$, and beaming, $\Omega$. The dash-dotted lines correspond to the BHNS model and the solid lines to the BNS model. The region shaded in grey represents the likely range of values for the primary neutron star's magnetic field. The space covered with constant $\epsilon_r=10^{-4}$ and range $0.01<\frac{\Omega}{4\pi}\leq 1$ is shaded in green. \textit{Right}: Total battery power generated in a inspiralling BHNS (equation \ref{eq:battery power}) and BNS (equation \ref{eq:battery_BNS_power}) system normalized to the power when the FRB is detected as a function of time from the FRB. We assume that the FRB is detected at the point of closest contact ($\frac{3GM}{c^2}$ for BHNS, 26\,km for BNS), denoted by the yellow star. We show the light curves of \frb\ and \frbb\,, where the peak corresponds to the FRB detection. The previous two time samples are denoted by points on the respective light curve. In both cases, the second previous sample is at the noise level of the respective dataset. We omit the portion of the light curve after the peak, as it is subject to propagation effects such as scattering, and is not representative of the intrinsic emission.}
    \label{fig:battery}
\end{figure*}

The most important unknown parameter in the battery emission mechanism model outlined in \textsection\ref{sect:battery} is the radio efficiency, $\epsilon_r$. Following equations \ref{eq:batteryBH} and \ref{eq:battery_BNS} for the BNS and BHNS inspiral models respectively, we plot the derived magnetic field $B$ of the primary neutron star as a function of $\frac{\epsilon_r}{\Omega}$ in the left-panel of Figure \ref{fig:battery}. The range of possible $B$ for neutron stars in such systems is uncertain but is thought to be $\sim10^{12}-10^{15}$\,G and is represented by the region shaded in grey. While the true radio efficiency is unknown, we use a fiducial value of $10^{-4}$ from pulsar studies \citep[see e.g.][]{Szary14} and a wide range of beaming values $0.01<\frac{\Omega}{4\pi}\leq 1$ (represented by the region shaded in green) for comparison. Generally though, the energetics of both FRBs fit this model for a wide range of parameter values. Fortunately, this model can be tested in another way. \citet{Mingarelli15} describe how a precursor to the main FRB may be detectable. The radio emission associated with this model is persistent, and increasing in luminosity with separation and time in a non-linear fashion. The luminosity surges at the time of coalescence and may account for the observed FRB. However given sufficient instrument sensitivity and resolution, the emission may be detected in earlier time samples at a fraction of the main FRB's signal-to-noise ratio (S/N) as a precursor. We can check whether a precursor to the main burst would have been detectable for \frb\ and \frbb\ by calculating the battery power (equation \ref{eq:battery power}) as a function of time ($r(t)\propto (t_{\textup{merger}} - t)^{1/4}$, for a circular orbit). We assume that the FRB is detected at the assumed point of closest contact ($\frac{3GM}{c^2}$ for BHNS and 26\,km for BNS). The result is shown in the right-panel of Figure \ref{fig:battery}. \frb\ was detected by ASKAP at 210 per cent of the detection threshold and the time resolution of the data is 0.864\,ms. The flux in the time sample that precedes the peak is about only 5 per cent lower. \frbb\ was detected at only 115 per cent of the detection threshold, and the sampling rate is $0.131$\,ms. The time sample preceding the peak is about half as bright. Comparing the relative intensities in the previous bins to what is expected from the models using Figure \ref{fig:battery}, we find that \frb's precursor bin is far too bright and the next prior sample far too faint. Even in the extreme case where coalescence occurs at $2R$, \frb\ still could not have been generated through this mechanism. As for \frbb, its light curve similarly rises far too rapidly. If we instead assume a spinning black hole for the BHNS case (where coalescence occurs at $\frac{GM}{c^2}$), the predicted rise in power is then too drastic. More generally, though, the shapes of the light curves do not match. Therefore we can exclude the battery model for both FRBs. The only caveat here is that we have assumed that the radio emission efficiency is constant over this few millisecond timescale.

\subsection{GRB jet model}
\begin{figure}
    \centering
    \includegraphics[scale=0.52]{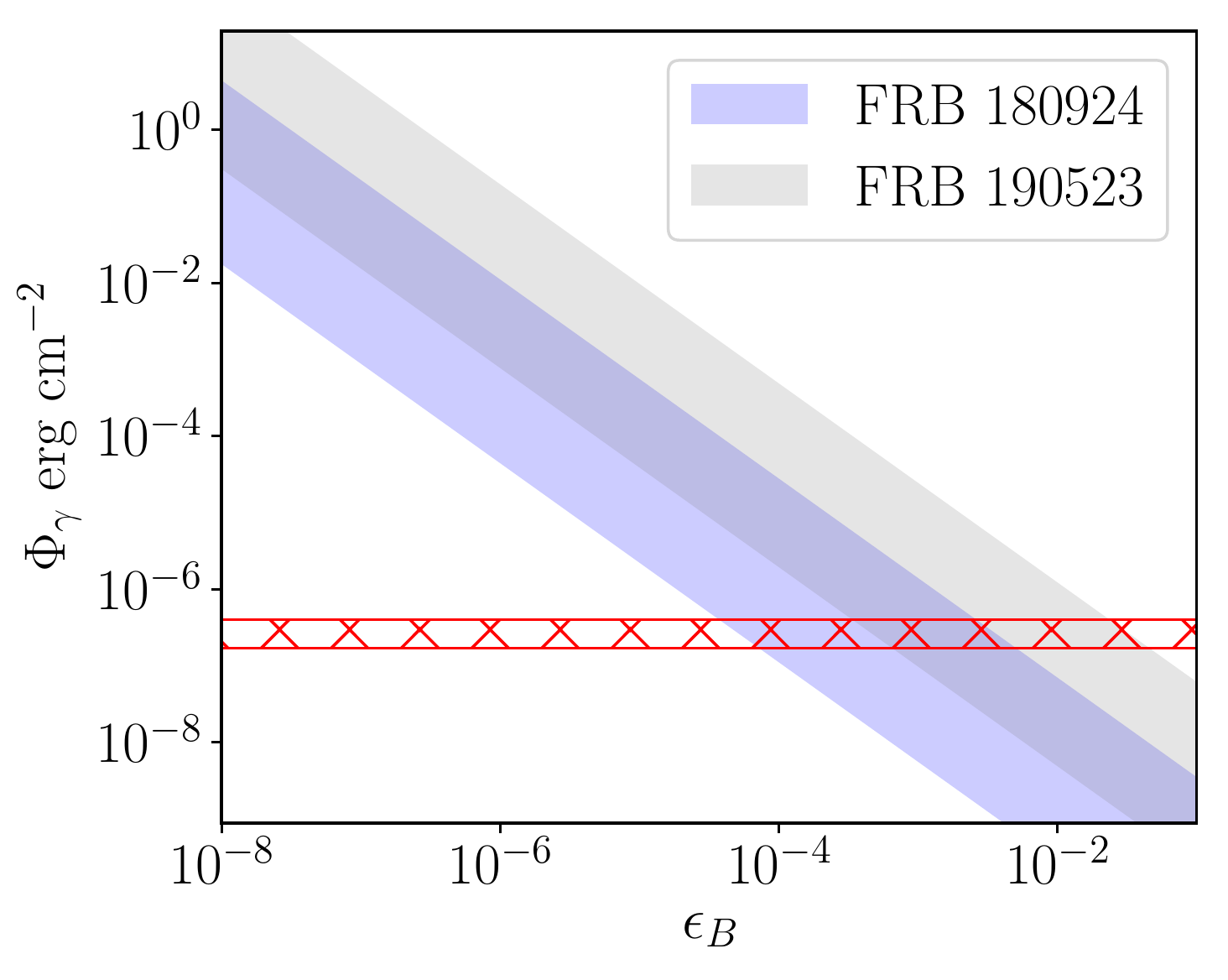}
    \caption{Following equations \ref{eq:usov_radio} and \ref{eq:usov_ratio}, expected GRB fluence ($\Phi_{\gamma}\propto \epsilon_B P_0^{-2}B$) as a function of fraction of wind energy contained within the magnetic field at the shock front ($\epsilon_B$), which is the biggest unknown parameter. The range of $\Phi_\gamma$ values corresponding to $10^{14}\,\text{G}\leq B \leq 10^{16}\,\text{G}$ and $0.001\,\text{s}\leq P_0 \leq 0.01\,\text{s}$ for both FRBs is shown by the corresponding shaded regions. The red hatched region denotes the \textit{Fermi} GBM and \textit{Insight}-HXMT upper limits on a GRB counterpart between 0.1\,s and 1\,s in duration for \frbb\ and \frb.}
    \label{fig:grb}
\end{figure}

The gamma-ray fluence, $\Phi_{\gamma}$, expected to accompany the FRB emission in this model scales with the three most uncertain quantities as $\epsilon_B P_0^{-2}B$. The expected gamma-ray fluence as a function of the fraction of wind energy contained in the magnetic field at the shock front ($\epsilon_B$) according to equations \ref{eq:usov_radio} and \ref{eq:usov_ratio} is shown in Figure \ref{fig:grb}. The resulting $\Phi_\gamma$ values are shown for a range of reasonable values of magnetic field $10^{14}\,\text{G}\leq B \leq 10^{16}\,\text{G}$ and initial spin $0.001\,\text{s}\leq P_0 \leq 0.01\,\text{s}$ \citep[e.g.][]{Rowlinson13}. The worst-case sensitivity of the \textit{Fermi} GBM for a 1-second burst is approximately $10^{-6}$\,erg cm$^{-2}$ (see \textsection\ref{sect:GRB search}). Therefore, for $\epsilon_B<10^{-4}$, a SGRB should have been detectable. Unfortunately the position of \frb\ was earth occulted when the FRB was emitted. However, \citet{Guidorzi2020} obtained upper limits in the 40--600\,keV band using \textit{Insight}-Hard X-ray Modulation Telescope (HXMT) data for various time integrations, including a $4.5\sigma$ upper limit of $4\times10^{-7}$\, erg cm$^{-2}$ for a 1-second duration GRB. \frbb\ was in the field of view of the \textit{Fermi} GBM and there is an upper limit of $4\times10^{-7}$\,erg cm$^{-2}$ for a 1-second duration GRB at the FRB's time and position (see \textsection\ref{sect:GRB search} and Figure \ref{fig:upperlimits}). These limits rule out the SGRB jet model with $\epsilon_B<4\times10^{-5}$ and $\epsilon_B<3\times10^{-4}$ for \frb\ and \frbb, respectively. We note that these results are particularly dependent on the spectral index of the radio emission, which we have assumed here to be $-1.6$. A much shallower spectrum could result in a lower GRB fluence that falls below the detection threshold of current GRB instruments. Additional joint gamma-ray and FRB datasets will be required to investigate this model further.

\subsection{Neutron star remnant}
\subsubsection{Rotational energy}
\label{sect:pulsar_disc}
\begin{figure}
    \centering
    \includegraphics[scale=0.55]{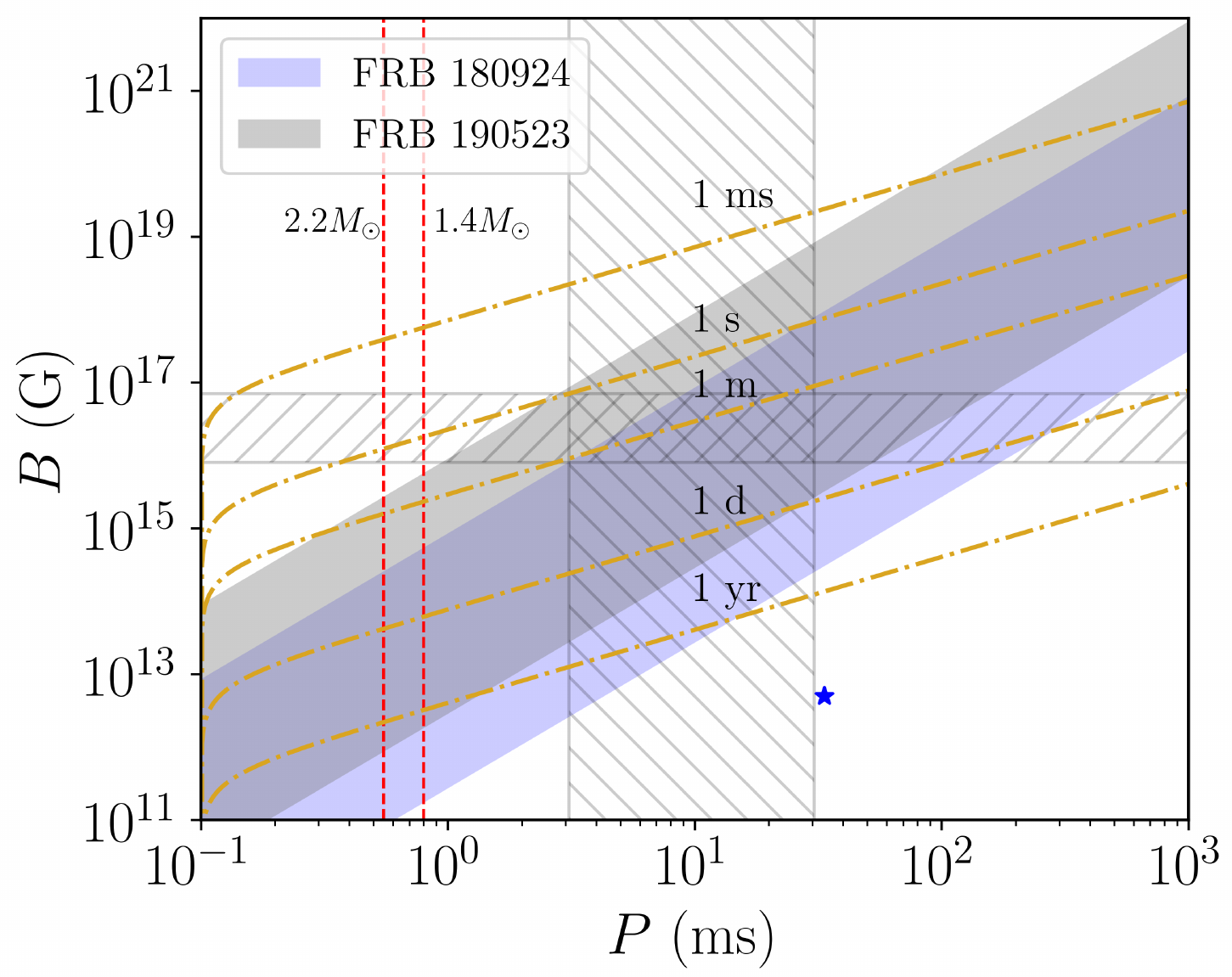}
    \caption{Magnetic field of a remnant pulsar emitting coherent radio emission following equation \ref{eq:pulsar}, as a function of pulsar spin period. The range of resulting values using $10^{-6}\leq \epsilon_r \leq 10^{-1}$ and $0.01 \leq \frac{\Omega}{4\pi} \leq 1$ for both FRBs is shown in the respective shaded regions. 
    The hatched regions represent the ranges of $B$ and $P$ for a typical neutron star  formed via a double neutron star merger, deduced from the X-ray plateaus of SGRBs. Lines of constant age are denoted by dashed yellow lines. Red vertical lines mark the theoretical neutron star breakup spin periods for two different masses. The Crab pulsar is represented by a blue star, for comparison.}
    \label{fig:pulsar}
\end{figure}

Following equations \ref{eq:pulsar} and \ref{eq:spindown} in \textsection\ref{sect:pulsar}, magnetic field strength of the supposed neutron star merger remnant as a function of pulsar spin period is represented in Figure \ref{fig:pulsar}. Thick shaded bands are used to show results for ranges of $10^{-6}\leq \epsilon_r \leq 10^{-1}$ and $0.01 \leq \frac{\Omega}{4\pi} \leq 1$, which are the dominant unknown variables. The results for $B$ are conservative because we have used the observing bandwidth to calculate intrinsic luminosity (equation \ref{eq:luminosity FRB}). Lines of constant neutron star age are shown for reference. An initial spin period of $0.1$\,ms has been assumed in order to show a wider parameter space, however such low values are not thought to be possible as they exceed the NS spin break-up values of $0.55$\,ms and $0.8$\,ms for neutron stars with a mass of $2.2$ \solar\ and $1.4$ \solar, respectively \citep{Lattimer04}.
If \frb\ and \frbb\ are produced according to this model, the NS would have to be very young, no more than a few years old for the former and no more than a few months for the latter.

The time window for parameters to be the right values in order to produce the observed FRB is particularly short for \frbb. After a spin-down time of about only one day, the model pushes the limits of the parameter space, requiring very high efficiency and narrow beaming. Therefore, within the realm of this model, it is likely that giant pulses, observed as FRBs, would only be emitted very shortly after the neutron star remnant is born. The pulsar subsequently spins down, its magnetic field decreases and the ingredients required to boost efficiency and/or beaming are no longer present or abundant enough to produce giant pulses detectable by radio telescopes on Earth. This is consistent with the lack of observed repeat bursts in follow-up observations of \frbb. The same case could be made for \frb. Another possible explanation could be that the neutron star remnant was unstable, collapsing soon after the FRB was produced into a black hole. A caveat to very early bursts (prior to a month post merger) are the effects of absorption that could obstruct any generated coherent emission. For longer time-scales of viability for this model, one might expect repeat bursts. However, the energy distribution of giant pulses spans several orders of magnitude and follows a power-law, with the brightest bursts being the least common \citep{Karuppusamy2010}. Given the fairly modest signal-to-noise ratios with which \frb\ and \frbb\ were detected, it is possible that fainter bursts are falling below the detection threshold.

Based on fits of the X-ray plateau of SGRBs, \citet{RowlinsonAnderson19} find that a typical magnetar remnant with detectable associated X-ray emission would have $B\sim10^{16}$\,G and spin period $\sim10$\,ms at birth. The precise range of derived values is included in Figure \ref{fig:pulsar}. An X-ray plateau from energy injection by a newborn NS would have been detectable for the range of $B$ and $P$ values that overlap the marked region. While simultaneous X-ray data of neither FRB are available, such datasets would constrain the properties of a remnant NS. The duration of X-ray plateaus from SGRBs has been observed to be as long as 3~hours, however most are less than $10$\, minutes \citep{Rowlinson13}. Considering the relatively short duration of X-ray plateaus, the target of opportunity observation latency is likely too long for instruments like \textit{Swift}/XRT \citep[minimum latency of 9 minutes and median 2~hours,][]{Burrows10}. Therefore, simultaneous radio and X-ray monitoring may be the only way to obtain a joint dataset. Alternatively, low-latency triggered radio observations following the detection of a GRB are also a possibility. An FRB search could then be conducted during the X-ray plateau phase that follows the detected GRB (previous such studies are \citealp{Bannister12,Obenberger14,Kaplan15,Anderson18,Rowlinson19}). Other possible avenues toward obtaining multi-wavelength and/or multi-messenger coverage of FRBs include triggered radio observations following the detection of neutron star mergers via their gravitational waves \citep{Yancey15,Abbott16,Kaplan16,Callister19}. 

\subsubsection{Magnetic energy}
\label{disc:mag}
We begin with the limit imposed on $B$ in the curvature radiation model from \textsection\ref{sect:curvature}. According to equation \ref{eq:curvature}, the magnetic field strength of the neutron star for \frb\ and \frbb\ respectively is approximately at least $3\times10^{12}$\,G and $1\times10^{15}$\,G. For increasingly beamed emission, this limit decreases.

There are several unknown and/or poorly constrained variables involved in deriving a predicted flux and emission frequency for the unstable synchrotron maser model outlined in \textsection\ref{sect:magnetar} and presented in \citet{lyubarsky2014maser} and elsewhere (e.g. \citealp{margalit18}, \citealp{metzger19} \& \citealp{Beloborodov17,Beloborodov19}). Many variables, though, are related to the nature of the upstream medium. For instance \citet{Beloborodov17} and \citet{metzger19} use electron-ions ejected from previous flares as the dominant material in which later ultra-relativistic ejections collide (as opposed to an electron-positron wind). Constraints on this model have been placed for \frb\ in \citet{metzger19}. If we instead assume that the nebula is powered by the spin-down wind of the magnetar  \citep{lyubarsky2014maser}, lower limits can be placed on the age of the magnetar based on the upper limits on persistent radio emission for each FRB and using equation \ref{eq:spindown} and the spin down age. We find a minimum age of $\sim8$\,months and $\sim1$\,week for \frb\ and \frbb, respectively.

\begin{figure}
     \centering
     \includegraphics[scale=0.55]{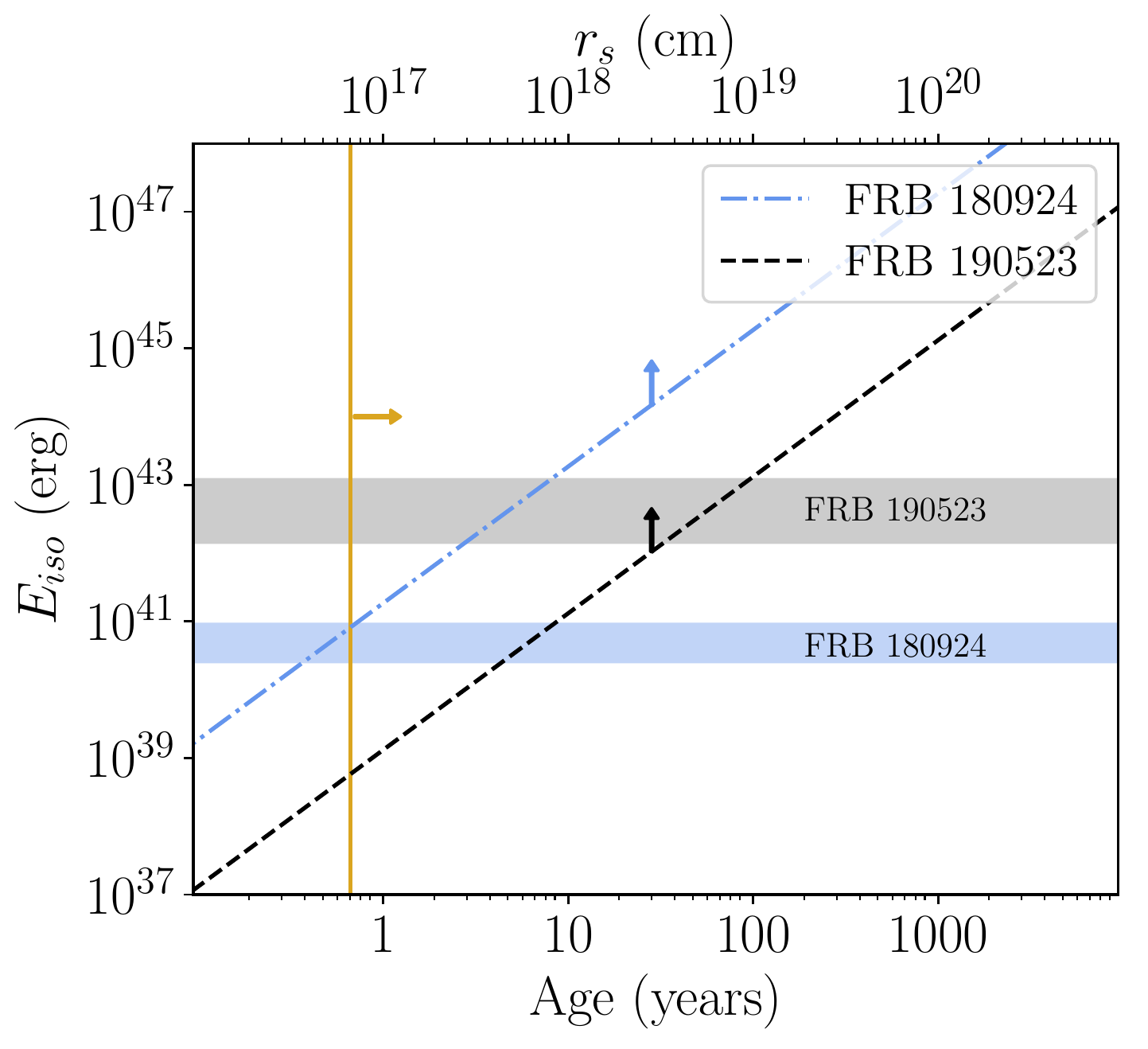}
     \caption{Lower limits on $E_{\text{iso}}$ of an FRB produced via the magnetar maser model outlined in \textsection\ref{sect:magnetar}, based on upper limits of persistent radio emission for \frb/\frbb\ are represented by the blue/black horizontal dash-dotted/dashed lines. The possible range of $E_{\text{iso}}$ according to equation \ref{eq:Eiso} using $\Delta \nu$ and $\nu_{\text{obs}}$ to obtain the minimum and maximum values, respectively, for each FRB is shown by the shaded regions. The vertical yellow line marks the minimum age (8 months) of the magnetar remnant associated with \frb\, based on limits of persistent nebular emission.}
     \label{fig:maser}
 \end{figure}

 According to equation \ref{eq:magnetar_energy_full}, the total isotropic emitted energy is proportional to unknown quantities as $E_{\text{iso}}\propto B \eta n b^2p^{-1}\xi^{-1}$. Estimates for $p$ can come from  measurements/upper-limits of persistent radio emission, assuming the FRB is produced in the nebula. Using the upper limits on the spin-down luminosity, $L_{\text{sd}}$, given by constraints on persistent radio emission, and using equation \ref{eq:pressure_balance}, we can obtain an upper limit on the pressure $p$ of the nebula ($p\propto L_{\text{sd}}r_s^{-2}$) assuming the distance, $r_s$, out to which the boundary between the nebula and wind occurs. A lower limit on $E_{\text{iso}}$ can then be placed using equation \ref{eq:magnetar_energy_full}, making assumptions for the remaining unknown variables. Using $b=0.01$, $B=10^{16}$\,G, $n=10^{-6}$\,cm$^{-3}$, and taking $\eta$ and $\xi$ to be the same value so that they cancel each other, we find:
\begin{equation}
    E_{\text{iso}} > 5.8 \times10^{41-45}\Bigg(\frac{\Delta t}{1\,\textup{ms}}\Bigg)\Bigg(\frac{L_{\text{sd}}}{10^{38}\,\textup{erg s}^{-1}}\Bigg)^{-1}\, \text{erg},\,\, r_s=10^{17-19}\,\text{cm}. \label{eq:E_iso_maser} 
\end{equation}
 This result is demonstrated in Figure \ref{fig:maser} for both FRBs. Also shown are calculations of $E_{\text{iso}}$ for both FRBs according to
\begin{equation}
    E_{\text{iso}} = 4\pi\mathcal{F} D^{2}\nu\,,
    \label{eq:Eiso}
\end{equation}
 where $\mathcal{F}$ is the measured burst fluence. The predicted spectrum of emission for this model is uncertain but thought to be complex \citep{Gallant92,PlotnikovSironi}. We therefore calculate a minimum and maximum value for $E_{\text{iso}}$ using $\Delta \nu$ and $\nu_{\text{obs}}$ respectively, in equation \ref{eq:Eiso}. Figure \ref{fig:maser} allows us to compare $E_{\text{iso}}$ derived from the model (equation \ref{eq:E_iso_maser}) to the values derived using equation \ref{eq:Eiso}. We use the following relationship for $r_s$ from \citet{Murase16}:
 \begin{equation}
 r_s \propto V_{\textup{ej}}^{3/5}P_0^{-2/5}M_{\textup{ej}^{-1/5}}T\,,
 \end{equation}
 where $V_{\textup{ej}}=0.2c$ is the merger ejecta velocity, $M_{\textup{ej}}=0.05$\,\solar\ is the ejecta mass, $P_0=10$\,ms is the initial spin of the magnetar remnant and $T$ is the age of the magnetar. 
 
 We find that, for our assumed model parameters, \frb\ could only have been produced by a magnetar flare shocking a nebula filled with an electron-positron plasma if its age, $T$, is $8\,\textup{months}<T<1$\,yr, else persistent emission would have been detected. The results for \frbb\ provide a larger range of ages, requiring $1\,\textup{week} < T < 100$\,yrs, however deeper radio searches for persistent emission are needed to provide more meaningful limits, as the current limit is 2 orders of magnitude weaker than that of \frb. We note that free-free absorption in the expanding merger ejecta can impede FRB propagation up to approximately one month post-merger \citep[e.g.][]{margalit19}. A flare with lower magnetic energy pushes the lower limit on the FRB energy down, whereas a denser nebula brings it proportionally higher.
 Constraints on the other variables of this model require a better theoretical understanding of magnetar flares and unstable synchrotron maser emission. We refer the reader to elaborate versions of the FRB maser emission theory treated in e.g. \citet{Beloborodov17,Beloborodov19,metzger19,PlotnikovSironi,Margalit19b} for more in-depth discussion and analysis on the unknown variables involved in this problem.
 
 
\subsubsection{NS collapse}
\begin{figure}
    \centering
    \includegraphics[scale=0.52]{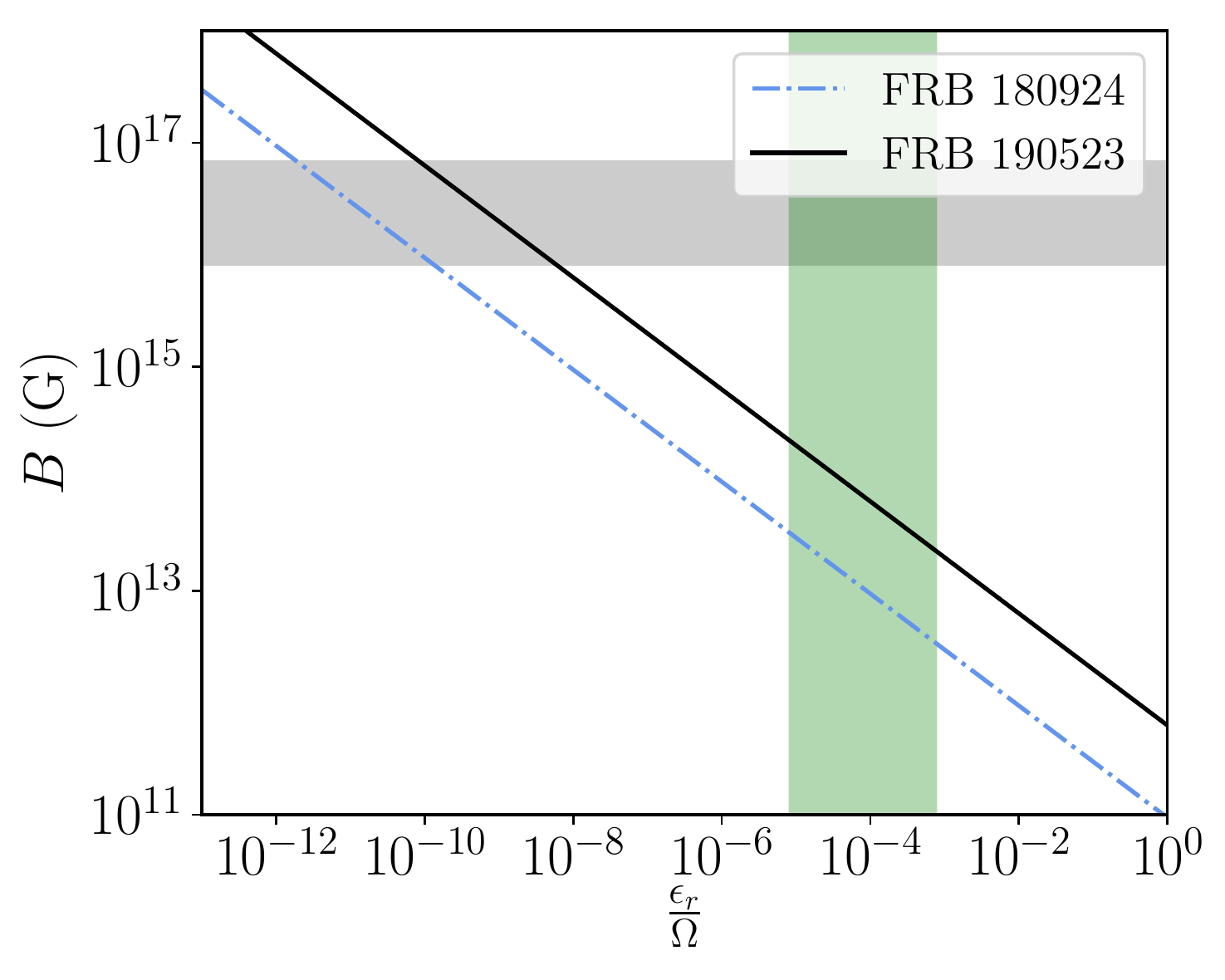}
    \caption{Derived magnetic field of remnant neutron star that collapses to produce observed FRB as a function of energy conversion efficiency and beam solid angle, according to equation \ref{eq:collapse}. The shaded horizontal grey band represents the range of $B$ expected for a newly born remnant neutron star with visible X-ray plateau. The shaded vertical green band denotes $\epsilon_r=10^{-4}$ for $0.01<\frac{\Omega}{4\pi}\leq 1$, shown for reference, though a wide range of values is acceptable.}
    \label{fig:collapse}
\end{figure}

Figure \ref{fig:collapse} shows the magnetic field of a remnant neutron star that collapses to produce the observed FRB as a function of energy conversion efficiency and beaming angle, according to equation \ref{eq:collapse} in \textsection\ref{sect:collapse}. As in Figure \ref{fig:pulsar}, the range of typical $B$ values for a neutron star remnant, based on X-ray plateau fits, is shown. The expected magnetic field of the neutron star depends on whether it is hypermassive (highly unstable) or supramassive (quasi-stable), and how long after formation the neutron star collapses. For instance, \citet{Piro19} find $10^{12}$\,G for a putative supramassive NS remnant of GW~170817. We use a fiducial energy conversion efficiency $\epsilon_r=10^{-4}$ as in Figure \ref{fig:battery} and a range of beaming angles $0.01<\frac{\Omega}{4\pi}<1$ to create the region shaded in green in Figure \ref{fig:collapse}. Our results show that if $\epsilon_r$ is comparable to that for pulsars, the magnetic field of the remnant neutron star must be 
$\sim10^{12-13}$\,G for \frb\ and $\sim10^{13-14}$\,G for \frbb. In this scenario, an X-ray plateau associated with the remnant prior to collapse would be too faint to detect \citep{Zhang01}:
\begin{equation}
L=1\times10^{45} \textup{erg s}^{-1}\Big(\frac{B}{10^{13}\,\textup{G}}\Big)^2\Big(\frac{P_0}{1\,\textup{ms}}\Big)^{-4}\Big(\frac{R}{10^6\,\textup{cm}}\Big)\,.
\end{equation}
 More generally, as one considers remnants with lower surface magnetic fields at the time of collapse, $\epsilon_r$ grows and increasingly narrower beaming is required. Ultimately, in order to provide better constraints on this model, multi-wavelength data are required. Given the non-detection of an SGRB in \textit{Fermi} GBM data in the 30 minutes preceding \frbb\ (\textsection\ref{sect:GRB} and Figure \ref{fig:upperlimits}) and the fact that most supramassive neutron stars collapse less than 10 minutes after their formation \citep{Rowlinson13}, it is unlikely (though possible) that the FRB is associated with the collapse of a short-lived neutron star formed post-merger. Alternatively, joint X-ray data could be used to probe the plateau emission that is expected to precede the collapse of the NS and FRB emission. As discussed in \textsection\ref{sect:pulsar_disc}, aside from simultaneous monitoring at both wavelengths, this would require rapid radio observations triggered by the detection of GRBs (both shorts and longs are relevant for this model, see for instance \citealp{Rowlinson19}).


\section{Conclusions}
\label{conclusion}
We have demonstrated how the information from localized FRBs can be utilized to test progenitor models. We have placed constraints on several emission models related to neutron star mergers and FRBs, for two recently localized sources, \frb\ and \frbb, which have environments reminiscent of the sites of neutron star mergers and SGRBs. We have ruled out the possibility of either FRB being produced during the final inspiral stages of a merging BNS or BHNS system through the interaction of the NS magnetosphere according to the unipolar-inductor model. We have performed a targeted sub-threshold search of \textit{Fermi} GBM data for a SGRB contemporaneous with either FRB, with no resulting promising candidates. We have demonstrated that either FRB could have been generated by a very young (less than one year old) remnant pulsar through rotational energy extraction, and that it would not have necessarily been accompanied by additional detectable bursts. We have shown that stringent limits on the age of a flaring magnetar with an electron-positron wind can be placed if deep observations constraining persistent radio emission are available. Fundamentally, all models used in this study depend on the magnetic energy density and the elusive method/efficiency of energy conversion (some version of $L\propto \epsilon_rB^2$). We have demonstrated the value of multi-wavelength datasets contemporaneous with FRB detections, which will ultimately be the best tool to break the degeneracy between possible models.  In particular, joint GRB/X-ray and FRB observations would provide meaningful constraints for many of the models presented here. For instance, while the energetics of both FRBs in this study are consistent with the collapsing neutron star model for a wide range of parameters, the non-detection of a SGRB counterpart renders the scenario less likely.

The number of localized FRBs is expected to increase drastically in the coming years, thanks to telescopes with the ability to localize single bursts to sub-arcsecond precision like ASKAP and the European VLBI Network. While we are limited in our ability to definitively reject or confirm some models presented in this work with only two FRBs, a larger sample will help move towards identifying their physical origin(s). To this end, we have laid out the ground work for future localized sources to be easily tested in the same way. We emphasize that all models except that in \textsection\ref{sect:battery} can be adapted to NSs born out of core-collapse supernovae (CCSN,the progenitors of LGRBs), for which the occurrence rate is much larger. The environment of CCSN is, however, denser and it may be difficult for any radio emission to escape shortly after the collapse occurs.

Finally, each of the models described in this work would have accompanying gravitational wave emission. Depending on the distance out to which an FRB is localized, sub-threshold GW searches can be conducted to provide further evidence for or against some of these models, for a given source. The next generation of gravitational wave detectors is expected to be 100 times more sensitive than the current instruments, which should suffice to confirm or reject these theories, if the origin of FRBs still remains unknown by then. 



\section*{Acknowledgements}
We thank A. Cooper, J. W. T. Hessels, P. Kumar, D. Stinebring and J. Weisberg for helpful discussions, and the referee for their constructive comments. A.G. acknowledges NASA funding through co-operative agreement NNM13AA43C.

\section*{Data availability}
The data used for the GRB search presented in this article are hosted through the Fermi Science Support Center and NASA's High Energy Astrophysics Science Archive Research Center. The data used for \frb\ and \frbb\ can be accessed at \url{https://heasarc.gsfc.nasa.gov/FTP/fermi/data/gbm/daily/2018/09/24/current/} and \url{https://heasarc.gsfc.nasa.gov/FTP/fermi/data/gbm/daily/2019/05/23/current/}, respectively.



\bibliographystyle{mnras}
\bibliography{references} 



\appendix
\section{The flaring magnetar model}
\label{sect:appendix}
Here, we show a detailed derivation of equation \ref{eq:magnetar_energy_full}, obtained following and building on the model presented in \citet{lyubarsky2014maser}. The magnetar flares start in the form of  magneto-hydrodynamic waves (Alfv\'{e}n waves) that propagate in the magnetosphere, sweeping up field lines to form a pulse that travels through the magnetar's wind. The magnetic field, $B_p$, stored in the pulse is some fraction, $b$, of the magnetic field at the magnetar's surface, $B$, and proportional to the magnetar's radius, $R$, and the pulse's distance from the magnetar surface, $r$:
\begin{equation}
    B_p = bB\frac{R}{r}, \quad b<1\,.
    \label{eq:b_pulse}
\end{equation}
The magnetar wind is composed  of magnetized electron positron plasma and its luminosity is determined by the spin-down luminosity $L_{\text{sd}}=\dot{E}$ defined in equation \ref{eq:spindown}. The wind's end boundary occurs when the wind's bulk pressure is balanced by the pressure confining the wind, $p$:
\begin{equation}
     p = \frac{L_{\text{sd}}}{4\pi r^2c}\,.
     \label{eq:pressure_balance}
\end{equation}
There is a termination shock at the radius at which this balance occurs, and a hot wind bubble (like a nebula) consequently forms. Therefore, $p$ is the pressure at the termination shock. 
The termination shock radius, $r_s$, is found by inserting equation \ref{eq:spindown} into equation \ref{eq:pressure_balance}:
\begin{equation}
    r_s = \sqrt{\frac{4\pi^3B^2R^6}{3pP^4c^4}}\,.
    \label{eq:termination}
\end{equation}

When the pulse reaches the termination shock, it meets a discontinuity as the upstream medium suddenly changes from the cold wind to the hot wind/nebula. It blasts the plasma in the nebula outward, generating a forward shock that propagates through the nebula's plasma. 
Equation \ref{eq:termination} can be substituted into equation \ref{eq:b_pulse} to find $B_p$ at the time of the blast:
\begin{equation}
    B_p = \frac{\sqrt{3} bp^{1/2}P^2c^2}{2\pi^{3/2}R^2}\,.
    \label{eq:b_pulse_solved}
\end{equation}
A contact discontinuity exists between the reverse and forward shocks, and defines a boundary for the shocked plasma in the nebula (think of the contact discontinuity moving with the propagating Alfven wave). At this contact discontinuity, the pressure (magnetic energy density, $\frac{B_p^2}{8\pi}$) of the pulse is equivalent to the bulk pressure of the hot plasma in the nebula crossing the forward shock. Since the pressure behind the shock is much greater than the unshocked plasma in the nebula ahead of the shock, we use the limiting density ratio which is 4 if we treat the plasma as a non-relativistic monatomic gas (adiabatic index $\gamma=5/3$) \citep{shocks}. Finally, we must consider that the contact discontinuity moves with Lorentz factor $\Gamma$ with respect to the observer. The particles in the plasma are boosted by a factor $\Gamma$ and the density too increases by $\Gamma$. The resulting pressure balance is then:
\begin{equation}
    \frac{B_p^2}{8\pi\Gamma^2} = 4\xi p \Gamma^2\,,\quad \xi<1\,,
    \label{eq:shock_balance}
\end{equation}
where dimensionless $\xi$ takes into account that some quantity of the high energy particles in the shocked plasma will lose their energy before they are able to enter the nebula, thereby decreasing the pressure. We solve for $\Gamma$ combining equations \ref{eq:b_pulse_solved} and \ref{eq:shock_balance}:
\begin{equation}
    \Gamma = \Big(\frac{3}{128}\Big)^{1/4}\frac{b^{1/2}cP}{\pi R \xi^{1/4}}.
\end{equation}

The magnetic field of the wind runs perpendicular to $B_p$ and the shock is mediated by that field. The gyration of the shocked particles creates an unstable synchrotron maser, that produces low-frequency emission, a fraction of which ($\eta$) manages to escape thermalization through the upstream unshocked plasma. For a pulse that travels a distance $\Delta r$ in the nebula, the isotropic energy of the escaped emission is ( \citealp{lyubarsky2014maser}, equation 11):
\begin{equation}
    E_{\text{iso}} = \eta 4\pi r_s^2 n m_e c^2\Gamma^2\Delta r, 
    \label{eq:magnetar_energy}
\end{equation}
where we have made use of the fact that $4\pi r_s^2 c n$ is the number of particles entering the shock per unit time and $n$ is the nebula's particle density. Finally, we use Doppler compression to find a relationship between observed burst duration $\Delta t$ and $\Delta r$ ($\Delta t = \frac{2\Delta r}{c\Gamma^2}$) and substitute $\Delta t$ into equation \ref{eq:magnetar_energy}, and, after full expansion, obtain:
\begin{equation}
    E_{\text{iso}} = \frac{\eta B^2R^2nm_ec^3b^2\Delta t}{16p\xi}.
\end{equation}

We now address emission frequency. The particles gyrate at the Larmor frequency
\begin{equation}
    \nu_p = \frac{eB_p}{2\pi m_ec\Gamma} = \Big(\frac{3}{2}\Big)^{1/4}\frac{b^{1/2}\xi^{1/4}ep^{1/2}P}{m_e\pi^{3/2}R}\,,
\end{equation}
where $m_e$ is the electron rest mass and $e$ is the electron charge. The radio emission is dominated by maser emission at this frequency \citep{lyubarsky2014maser}. The value of $\nu_p$ ranges from tens to hundreds of megahertz depending mostly on the pressure of the nebula $p$ ($P$ for a magnetar is likely to be approximately one second). However, for magnetically dominated plasmas, particle-in-cell simulations reveal complex shock structure that actually increases the peak frequency by several factors (for high magnitization, $
\sigma>1$,) \citep{PlotnikovSironi}. Furthermore, the spectrum of emission extends to higher frequencies \citep{Gallant92,PlotnikovSironi}.  In this way, GHz frequencies can be attained.


\bsp	
\label{lastpage}
\end{document}